\documentclass[twocolumn,
    showpacs,
    preprintnumbers,
    amsmath,
    amssymb,
    nofootinbib,
    longbibliography,
    tightenlines,
    nobibnotes,
    prb]{revtex4-1}
\usepackage{amsfonts}
\usepackage[english]{babel}
\usepackage[utf8]{inputenc}
\usepackage[T1]{fontenc}
\usepackage{times}
\usepackage{mathrsfs}
\usepackage{graphicx}
\usepackage{dcolumn}
\usepackage{bm}
\usepackage[colorlinks,bookmarks=true,citecolor=blue,linkcolor=red,urlcolor=blue]{hyperref}
\usepackage[tight, FIGTOPCAP, hang, raggedright, nooneline]{subfigure}

\usepackage{braket}

\renewcommand{\vec}[1]{\boldsymbol{\mathbf{#1}}}

\usepackage{xcolor}

\newcommand{\pd}[2]{\frac{\displaystyle \partial #1}{\displaystyle\partial #2}} 

\renewcommand{\d}[1]{\textrm{d}#1}

\begin{document}

\title{Quantum Oscillations and Magnetoresistance in Type-II Weyl Semimetals $-$ \\Effect of a Field Induced Charge Density Wave}

\author{Maximilian Trescher,$^1$  Emil J. Bergholtz$^{2}$ and Johannes Knolle$^{3}$ }
\affiliation{$^1$ Dahlem Center for Complex Quantum Systems and Institut f\"ur Theoretische Physik, Freie Universit\"at Berlin, Arnimallee 14, 14195 Berlin, Germany
\\ $^2$ Department of Physics, Stockholm University, AlbaNova University Center, 106 91 Stockholm, Sweden
\\ $^3$ Blackett Laboratory, Imperial College London, London SW7 2AZ, United Kingdom} 
\date{\today}

\begin{abstract}
Recent experiments on type-II Weyl semimetals such as WTe$_2$, MoTe$_2$, Mo$_x$W$_{1-x}$Te$_2$ and WP$_2$ reveal remarkable transport properties in presence of a strong magnetic field, including an extremely large magnetoresistance and an unusual temperature dependence. Here, we investigate magnetotransport via the Kubo formula in a minimal model of a type-II Weyl semimetal taking into account the effect of a charge density wave (CDW) transition, which can arise even at weak coupling in the presence of a strong magnetic field because of the special Landau level dispersion of type-II Weyl systems. Consistent with experimental measurements we find an extremely large magnetoresistance with close to $B^2$ scaling at particle-hole compensation, while in the extreme quantum limit there is a transition to a qualitatively new scaling with approximately $B^{0.75}$. We also investigate the Shubnikov-de Haas effect and find that the amplitude of the resistivity quantum oscillations are greatly enhanced below the CDW transition temperature which is accompanied by an unusual non-monotonous (non-Lifshitz-Kosevich) temperature dependence. 
\end{abstract}

\maketitle

\section{Introduction}
Condensed matter incarnations of Weyl fermions, come in two distinct flavours \cite{armitage_weyl_2018}: type-I, which are direct analogues of their high-energy cousins\cite{weyl_elektron_1929}, occur in true semimetals with a point-like Fermi surface\cite{volovik_universe_2009,murakami_phase_2007,wan_topological_2011,hosur_recent_2013,xu_discovery-taas_2015,lv_discovery_2015}, and type-II where the linear dispersion is so strongly tilted that Fermi-pockets are formed around the Weyl points\cite{bergholtz_topology_2015,soluyanov_type-ii_2015}. Immediately following their discovery, it was realised that type-I Weyl fermions were associated with novel transport phenomena such as the chiral anomaly\cite{nielsen_adler-bell-jackiw_1983,liu_chiral_2013,son_berry_2012,grushin_consequences_2012,zyuzin_topological_2012,goswami_axionic_2013,parameswaran_probing_2014,klier_transversal_2015,klier_transversal_2017,behrends_transversal_2018,huang_observation_2015,zhang_signatures_2016}. In contrast, due to the concealed nature of the type-II Weyl fermions it is a priori less obvious that they would lead to interesting measurable effects. However, transport measurements on type-II Weyl semimetals have arguably proven even more intriguing: most saliently, experiments on type-II Weyl semimetals found an extremely large magnetoresistance that did not saturate until the highest accessible magnetic fields\cite{ali_large-magnetoresistance-wte2_2014,kumar_extremely_2017,zeng_compensated_2016}. In all these experiments a scaling of the magnetoresistance close to $B^2$ was found. Furthermore, the magnetotransport in these materials is strongly dependent on the direction of the applied magnetic field. 

\begin{figure}[h]
    \centering
    \includegraphics[width=0.47\textwidth]{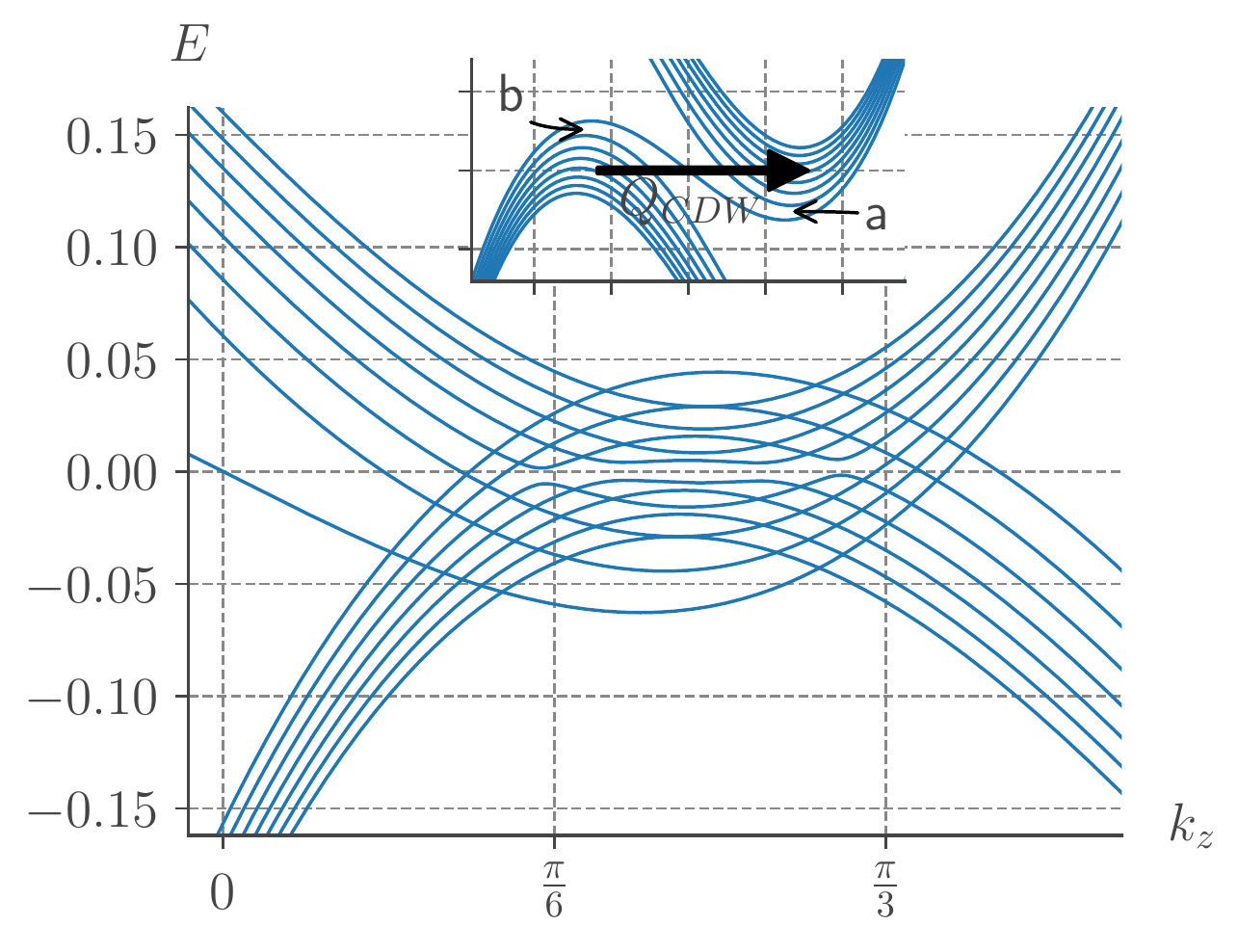}
    \caption{Dispersion of a type-II Weyl semimetal with a magnetic field ($B=7.5T$) induced CDW as given in \eqref{ll-dispersion}. Energy is given in units of $eV$, $k_z$ in reciprocal lattice vectors. 
    The main panel shows the folded dispersion in the presence of a CDW, which gaps out part of the Landau levels, with the lower branch ($b$) shifted by $Q_{CDW}$.
    The inset highlights the field induced nesting property  between electron and hole type pockets of the normal state with the CDW wave vector $Q_{CDW}$.
    It further includes labels for the two branches $a$ and $b$.
    }
    \label{fig:weyl2-ll-dispersion}
\end{figure}

In addition, a number of experiments found an unusual temperature dependence of transport properties\cite{thoutam_temperature-dependent_2015,wu_temperature-induced_2015,fatemi_magnetoresistance_2017,ramshaw_unmasking_2017}. On the one hand, these could be related to the strong temperature dependence of the chemical potential, as expected for a low density system when $k_B T$  and the effective Fermi energy are of similar order of magnitude. On the other hand, it could be a direct signature of a many-body instability, such as a field induced excitonic insulator\cite{khveshchenko_magnetic-field-induced_2001} or a density wave transition\cite{trescher_charge_2017}. The CDW scenario was put forward in a recent work of us\cite{trescher_charge_2017} motivated by the observation that the Landau level structure of type-II Weyl semimetals\cite{udagawa_field-selective_2016} shows emergent nesting properties as illustrated in Fig. \ref{fig:weyl2-ll-dispersion} (inset). At zero field, the electron and hole pockets have different shapes. However, a strong magnetic field leads to nested, quasi-one-dimensional electron and hole pockets prone to a CDW instability already for weak interactions. Such an intra-Weyl cone CDW would have a small wave vector and the associated breaking of translational symmetry should be observable in scattering experiments. Here, we show that a field induced CDW leads to a very rich phenomenology of magnetoresistance properties in type-II Weyl semimetals.  

\section{Theory}
We consider a generic and minimal model of a type-II Weyl semimetal, which can be theoretically derived as a low energy expansion of a lattice model\cite{udagawa_field-selective_2016} as given in Ref. \onlinecite{trescher_charge_2017}. A crucial observation there was to
include not only terms linear in momentum but also the next non-zero terms (in this case $k_z^3$) to account for electron and hole pockets, see Fig.\ref{fig:weyl2-ll-dispersion}.

We use the general low energy theory of a type-II Weyl semimetal with additional third-order terms in $k_z$ along the tilt direction and concentrate on one of the Weyl cones
\begin{align}
     \hat H_0^{\text{eff}} \,= &  (-\eta k_z + \gamma k_z^{3}) \sigma_0 + (k_z + \beta k_z^3) \sigma_z +
      k_x\sigma^x + k_y\sigma^y .
\label{Eff_Hamiltonian}
\end{align}
The parameters $\eta = -(t_1 + 2t_2),  \beta = -\frac{1}{6}$ and $\gamma = -\frac{1}{6} (t_1 + 8t_2)$ are directly related to the lattice model's hopping parameters, where we choose $t_1=-0.8$ and $t_2=-0.6$.
Throughout this work, for concreteness, we consider the exemplary values $\hbar v_F = 4 \,\textrm{eV\AA}$ and the lattice constant of $a_0 = 28 \,\textrm{\AA}$. 
Note, in this simple tight binding model the electron and hole pockets extend to a large fraction of the first Brillouin zone (BZ),
whereas in typical materials they only cover a much smaller fraction.
Here the size of the BZ (the corresponding inverse unit cell) is not important, as the size of the electron and hole pockets defines the relevant length scales.
In consequence we chose a relatively large lattice constant for the exemplary model in order to obtain small electron and hole pockets in momentum space,
so that they are similar to the ones in actual materials.

For simplicity and transparency, we focus on situations with the magnetic field directed along the tilt of the Weyl cone, i.e. $B \parallel z$ when the tilt direction is along $k_z$, and only briefly comment on the generalization to generic directions in the discussion.
The field $\vec{B} = \nabla \times \vec{A}$ is then introduced by a minimal coupling of the vector potential, which we consider to be $\vec{A} = \left( 0, Bx, 0 \right)$ in the Landau gauge, to the crystal momentum $\vec{\Pi} = \hbar \vec{k} - \frac{e}{c}\vec{A}$.
We define the standard raising and lowering operators $a = \frac{l_B}{\sqrt{2} \hbar} \left( \Pi_x - i \Pi_y \right)$ and $a^\dagger = \frac{l_B}{\sqrt{2}\hbar} \left(\Pi_x + i \Pi_y\right)$ with $[a, a^\dagger] = 1$,
where we introduced the magnetic length $l_B = \sqrt{\frac{\hbar}{e B}}$.
Consequently, we arrive at the following Hamiltonian describing the low energy theory
\begin{widetext}
\begin{align}
    H_0 = \int \d^3 r 
    \begin{pmatrix}
        \Psi_A^\dagger(\vec{r})
         & \Psi_B^\dagger(\vec{r})
    \end{pmatrix}
    \hbar v_F
    \begin{pmatrix}
        (1-\eta)k_z + (\gamma + \beta) k_z^3 & \frac{\sqrt{2}}{l_B} \hat{a}^\dagger \\
        \frac{\sqrt{2}}{l_B} \hat{a} & -(1+\eta)k_z + (\gamma - \beta) k_z^3
    \end{pmatrix}
    \begin{pmatrix}
        \Psi_A(\vec{r}) \\
         \Psi_B(\vec{r})
    \end{pmatrix}
    \label{weyl2-ll-hamiltonian}
\end{align}
\end{widetext}
with the wave functions
\begin{align}
\Psi_A^\dagger(\vec{r}) &= \sum_{n,p,k_z} e^{i k_z z} \psi_{n,p}(x,y) \hat{c}_{A, n, p, k_z} \\ 
\Psi_B^\dagger(\vec{r}) &= \sum_{n,p,k_z} e^{i k_z z} \psi_{n-1,p}(x,y) \hat{c}_{B, n, p, k_z}
\label{wavefunctions}
\end{align}
and the normalized Harmonic Oscillator wave functions $\psi_{n,p}(x,y)= \frac{1}{\sqrt{L}} e^{i p x } \left( \pi 2^{2n} (n!)^2 \right)^{-1/4} e^{-\frac{1}{2} (y+p)^2} H_n(y+p)$  including the Hermite polynomials $H_n$. Note, that there is an extra shift of the Landau level index from $n$ to $n-1$ for the $B$ sublattice part of the wave function. 

In the braket notation we write the states as spinors in the $A/B$ basis and perform a  rotation such that the Hamiltonian becomes diagonal
\begin{align}
\begin{pmatrix}
\Ket{n,k_z,p,a} \\
\Ket{n-1,k_z,p,b}
\end{pmatrix}
    &= U(n,k_z,p)
    \begin{pmatrix} 
        \Ket{n,k_z,p,A} \\
        \Ket{n-1,k_z,p,B} 
\end{pmatrix}
\label{basistrafo}
\end{align}
where $U$ is a unitary transformation that is easily obtained from \eqref{weyl2-ll-hamiltonian}.  
The energies, labeled by $s\in a,b$ for  $\Ket{n,p,k_z,s}$ are given by 
\begin{align}
    E_{n,a/b}(k_z) &= \hbar v_F \left(  -\eta k_z  + \gamma k_z^3 \pm \sqrt{(k_z + \beta k_z^3)^2 + \frac{2}{l_B^2}|n|} \right)
    \label{ll-dispersion}
\end{align}
for $n\neq 0$. For $n=0$, the energy of the chiral level is
\begin{align}
    E_{0,k_z} &= \hbar v_F \left((1 - \eta) k_z + (\beta + \gamma) k_z^3 \right).
\end{align}
The Landau level degeneracy becomes apparent as $E$ does not depend on $p$.
Note that the chiral mode corresponds to the $(1 \; 0)^T$ spinor already in the original basis.
Therefore, there is no basis transformation $U(n=0)$ for the chiral level.

Following our recent work Ref. \cite{trescher_charge_2017}, we take into account the effect of an interaction induced CDW. Due to the multi-band nature of the Landau level dispersion it is sufficient to restrict the analysis of the weak coupling instability to the simplest form of a  contact interaction. 
This interaction, projected onto the Landau level bands, is given by
\begin{widetext}
\begin{align}
\label{ContInt} 
H_{\textrm{int}}\! =\! 
    \frac{U}{2}  \!
  \sum_{\substack{n_1,n_2,n_3,n_4,\\p_1,p_2,k_z,k_z',\\q_x,q_y,q_z}} \! \!
     e^{iq_y ( p_1 - p_2 - q_x)} 
         J_{n_4,n_1}(\mathbf{q}) J_{n_3,n_2}(-\mathbf{q}) \!\!
    \sum_{\alpha,\beta=A,B} \!
     c^{\dagger}_{\alpha,n_1,p_1,k_z} c^{\dagger}_{\beta,n_2,p_2,k_z'} 
     c_{\beta,n_3,p_2+q_x,k_z'+q_z} c_{\alpha,n_4,p_1-q_x,k_z-q_z} \, .
\end{align}
\end{widetext}
Details of the matrix elements $J(q)$ and useful analytic simplifications of them are given in Appendix \ref{sec:matrixelements} for our choice of a uni-directional CDW with $q_x = q_y =0$.

At strong magnetic fields, even weak interactions lead to CDW transitions because of the quasi-nesting between the electron and hole pockets, $E_{n,a} (k_z)\approx -E_{n,b}(k_z+Q_{CDW})$ (see Fig.\ref{fig:weyl2-ll-dispersion}). 
In a mean-field framework, as presented in Ref. \onlinecite{trescher_charge_2017}, the quartic interactions are decoupled, leading to an effective CDW Hamiltonian
\begin{align}
    & H_{\textrm{MF}, n}\left( k_z \right) = \nonumber\\
    & \begin{pmatrix}\!
        a^{\dagger}_{n,k_z} & b^{\dagger}_{n,k_z - Q}
    \!\end{pmatrix}
        \begin{pmatrix}\!
            E_a(n, k_z) & P(k_z) \\
            P(k_z)^* & E_b(n, k_z-Q)
        \!\end{pmatrix}
    \begin{pmatrix}\!
        a_{n,k_z} \\ b_{n,k_z-Q}
    \!\end{pmatrix} .
    \label{hmf}
\end{align}
The analytical form of the off-diagonal matrix elements $P$ is derived from the contact interaction projected to the Landau level band structure and given in Ref. \onlinecite{trescher_charge_2017} and the appendix \ref{sec:matrixelements}.

\subsection{Kubo formula of the conductivity}
We concentrate on the transport properties perpendicular to the magnetic field, which are obtained from the in-plane conductivity $\sigma_{\alpha \alpha}$ with $\alpha=x,y$.
We apply the standard Kubo formula for conductivity, which is given by\cite{tabert_particle-hole_2015,shao_magneto-optical_2016}:
\begin{align}
    &\sigma_{\alpha \beta}(\omega)=\nonumber\\ &\frac{i\hbar}{2 \pi l_B^2} \sum_{\zeta,\zeta'}
    \frac{f(E_{\zeta'}) - f(E_{\zeta}) }{ E_{\zeta} - E_{\zeta'}}
    \frac{ \Braket{\zeta | \hat{j}_\alpha | \zeta'} \Braket{\zeta' | \hat{j}_\beta | \zeta} }
    {\hbar \omega + E_{\zeta'} - E_{\zeta} + i\hbar / (2\tau)}
    \label{kuboLL}
\end{align}
using the short-hand notation $\zeta = \{n, k_z, a/b\}$, $f(E)$ for the Fermi function, and $\tau$ is a scattering-induced lifetime.
The summation over $p$ is already executed and accounts for the Landau level degeneracy factor $\frac{1}{2\pi l^2_B}$. We assume a constant time $\tau$ due to impurity scattering. Note, this simplifying assumption is motivated from the very weak frequency dependence of the density of states (DOS) of a type-II Weyl semi-metal with finite electron and hole pockets. For example in Weyl type-I systems with a vanishing DOS a more elaborate calculation of the scattering time is necessary\cite{klier_transversal_2015}.

To evaluate the Kubo formula we calculate the current operators as
\begin{align}
    \hat{j}_\alpha &= e \hat{v}_\alpha = \frac{e}{\hbar} \pd{\hat{H}}{\Pi_\alpha}
    \label{currentoperator-general}
\end{align}
with the Hamiltonian matrix, Eq.\eqref{weyl2-ll-hamiltonian}, in the form
\begin{align}
    H &=
    \hbar v_F
    \begin{pmatrix}
        (1-\eta)k_z + (\gamma + \beta) k_z^3 & \frac{1}{\hbar} (\Pi_x + i \Pi_y)  \\
        \frac{1}{\hbar} (\Pi_x - i\Pi_y) & -(1+\eta)k_z + (\gamma - \beta) k_z^3
    \end{pmatrix} \, ,
    \label{}
\end{align}
so 
\begin{align}
    \hat{j}_x &= e v_F \sigma_x \\
    \hat{j}_y &= - e v_F \sigma_y
    \label{currentoperator}
\end{align}
From this, we obtain the relevant matrix elements in the basis of Eq.\eqref{weyl2-ll-hamiltonian}
\begin{align}
    \Braket{n,k_z|\hat{j}_x|n',k_z'}_{A/B} = \\ \nonumber
    \Braket{n,k_z|
    \begin{pmatrix}
        0 & \delta_{n,n'-1} \\
        \delta_{n-1,n'} & 0 
    \end{pmatrix}
    |n',k_z'}_{A/B} \,.
\end{align}
These current operators are then transformed with two subsequent unitary transformations, first with Eq.\eqref{basistrafo} and second with the one that diagonalizes the CDW  Hamiltonian Eq.\eqref{hmf} with the mean-field energies $E_{\zeta}$ appearing in the Kubo formula.

\begin{figure}[ht]
    \centering
    \includegraphics[width=0.47\textwidth]{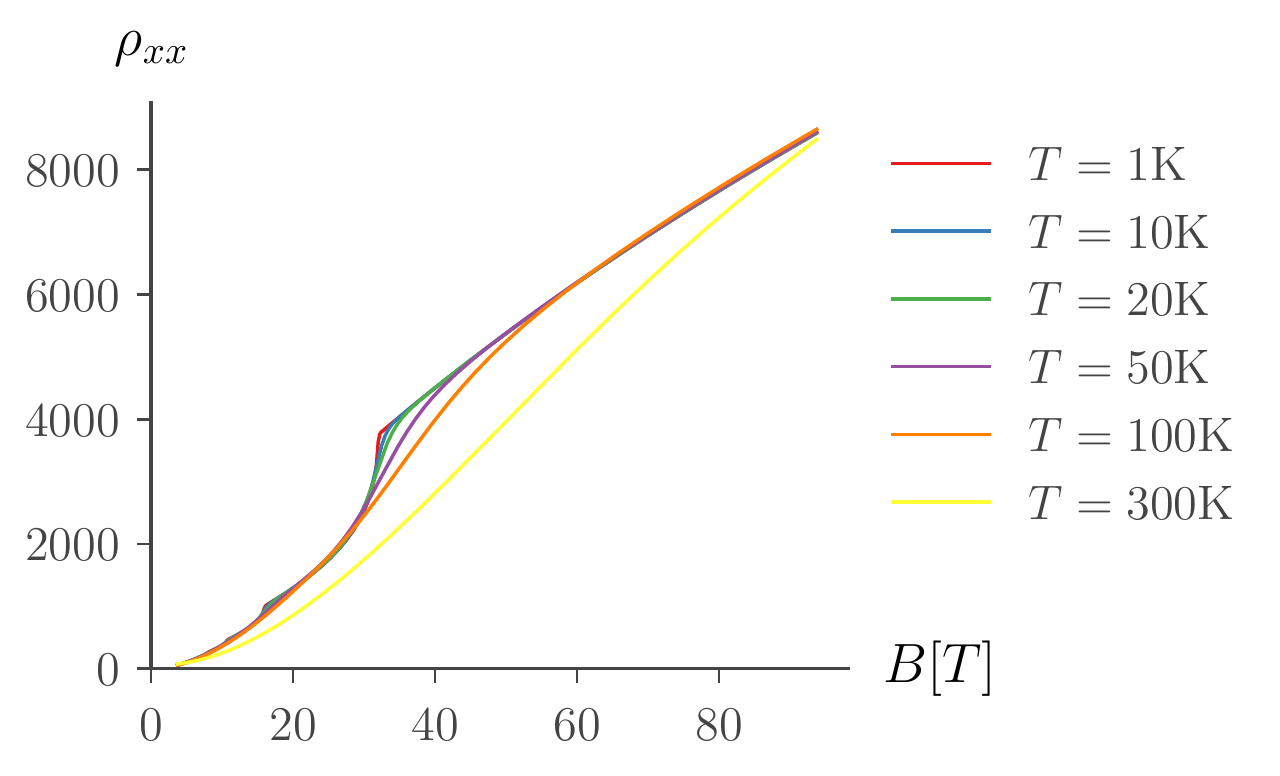}
    \includegraphics[width=0.47\textwidth]{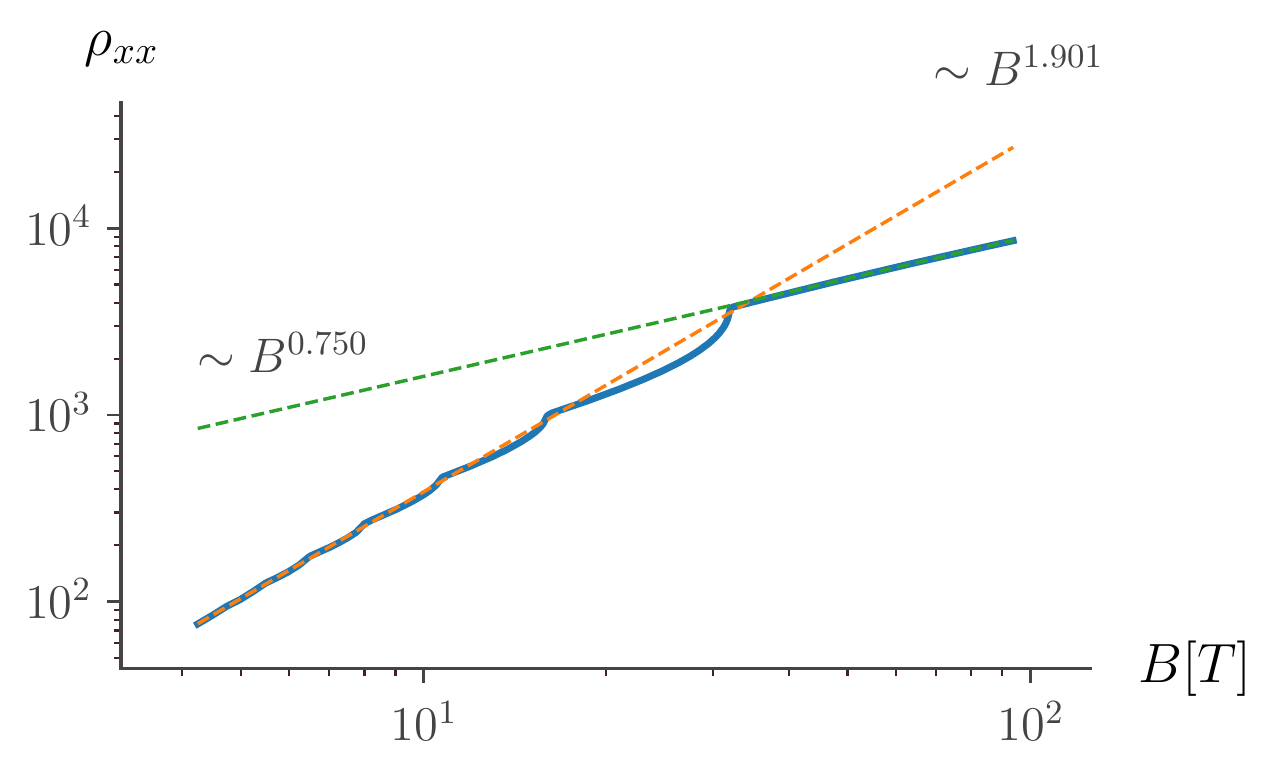}
    \caption{Transversal resistivity $\rho_{xx}$ (arbitrary units) vs. magnetic field (in Tesla) for different temperatures $T$. 
        The top panel shows clearly the crossover around $B_Q \simeq 31 \textrm{T}$ and additionally pronounced quantum oscillations (at low temperatures and $B < B_Q$).
    The bottom panel displays the $T=1 K$ data of the top panel on a log-log scale.
    The dashed lines show least-squares fits for different regimes of B. 
    The corresponding exponents for the fitted power law behaviour are indicated.}
    \label{fig:clean-t-dependence}
\end{figure}

\section{Results}
The magnetoresistance is usually given in $\%$ and defined as
\begin{align}
    MR(B) = \frac{\rho(B) - \rho(0) }{\rho(0)} \, .
    \label{magnetoresistance}
\end{align}
There is no CDW at zero magnetic field, therefore $\rho(0)$ in our treatment is equal in the cases with and without interactions.
We are mainly interested in the qualitative behaviour of the magnetoresistance and not in exact quantitative values, which would require a detailed microscopic description of a specific material.
Hence, we concentrate on the behavior of the resistivity $\rho{(B)}$, which is qualitatively equivalent to the magnetoresistance.
We consider the case with zero chemical potential where the dispersion relation is particle-hole symmetric. 

A hypothetical imbalance between particle and hole pockets leads quickly to a saturation of MR in the semiclassical picture.
We verified that this predictions holds also for the quantum mechanical computation for the non-interacting case with a constant scattering time. An extension to the CDW case would require further adaptations to the mean-field calculations beyond the scope of the present work.

\subsection{Magnetoresistance}
In the compensated situation considered here, the Hall conductivity $\sigma_{xy}$ vanishes and $\sigma_{xx} = \sigma_{yy}$ because of the rotational symmetry.
Hence, the resistivity along the in-plane $x$-direction can be directly calculated from the corresponding conductivity
\begin{align}
    \rho_{xx} &= \frac{\sigma_{xx}}{\sigma_{xx}^2 + \sigma_{xy}^2} = \frac{1}{\sigma_{xx}} \,.
    \label{rho_simple}
\end{align}

First, we discuss the case without interactions and hence without CDW.
The full temperature dependence as calculated from the Kubo formula Eq.\eqref{kuboLL} is shown in Fig.\ref{fig:clean-t-dependence}.
For small fields, our results agree well with a semiclassical model of two bands with perfectly compensated particle and hole pockets\cite{ashcroft_solid_1976} predicting a scaling of the resistivity as $\sim B^2$. 
Note that the numerical fit of the data yields powers of $B$ slightly smaller than 2 because the oscillations at higher field are not symmetric around the expected $B^2$ background.
Therefore, a numerical fit will weakly depend on the range of magnetic fields considered.
At the same time we note that several experiments\cite{kumar_extremely_2017,ali_large-magnetoresistance-wte2_2014,zeng_compensated_2016} experimentally found a scaling with powers between $1.8$ and $2$, similar to our results.

The semiclassical prediction obviously neglects the quantum oscillations due to the discreteness of the Landau levels, which become more pronounced at higher magnetic fields and lower temperature.
We define the magnetic field strength $B_Q$ as the minimum field at which the lowest Landau level reaches the Fermi energy. Beyond this \textit{quantum limit} $B>B_Q$ only the chiral mode contributes to the density of states at zero energy. In addition to the semiclassical quadratic scaling for $B \ll B_Q$
we observe a crossover to a different scaling $\rho \sim B^{0.75}$ above the quantum limit, see Fig.\ref{fig:clean-t-dependence} (bottom panel).

The magnetic field strength associated with the crossover to the quantum limit is $B_Q \simeq 31 T$ for our exemplary tight binding model.
This magnetic field strength $B_Q$ depends strongly on the size of the electron and hole pockets.
We denote the size of the pocket with $k_P$.
The corresponding length scale is given by $1/k_P$.
As the magnetic length $l_{B_Q}$ at the quantum limit is proportional to that length scale $1/k_P$,
we find $B_Q \propto k_P^2$.

\begin{figure}[ht]
    \centering
    \includegraphics[width=0.48\textwidth]{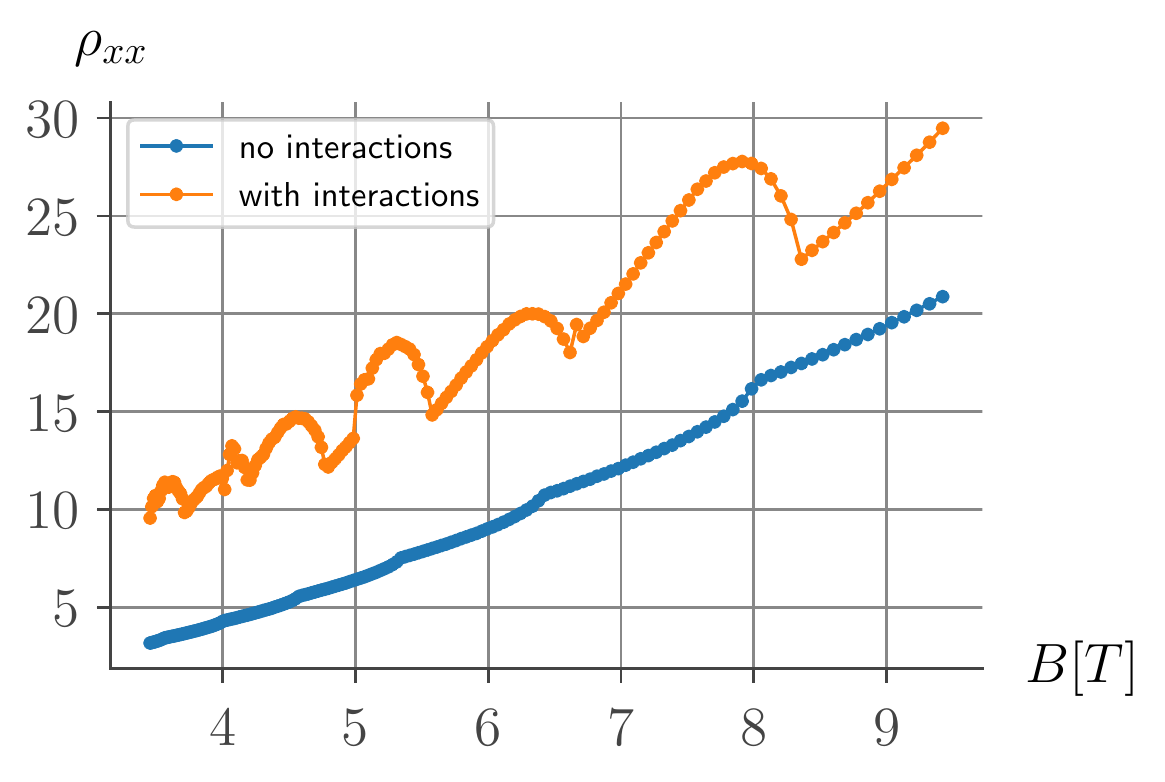}
    \caption{Transversal resistivity $\rho_{xx}$ (arbitrary units) for a Weyl semimetal without interactions and with an interaction induced CDW at a temperature of $T = 0.5 K$.}
    \label{fig:rho-cdw}
\end{figure}
Next, we include the effect of a low temperature CDW transition. 
We will use the same exemplary parameters as in our previous work \cite{trescher_charge_2017}.
There, we observed that at low temperatures a constant fraction ($\simeq 0.5$) of the Landau levels crossing the Fermi energy is gapped out by the CDW, regardless of the magnetic field.
This statement is only valid in the regime  $B \ll B_Q$ studied in Ref.~\onlinecite{trescher_charge_2017}, where the fraction is nearly a continuous number, due to the large number of Landau levels, while for higher fields we expect this fraction to vary due to the discrete number of Landau levels.
In the quantum limit there is no Landau level left that could be gapped by a CDW, therefore the fraction of Landau levels gapped by the CDW then drops to zero.

The resulting resistivity $\rho_{xx}$ obtained from the mean-field results is shown in Fig.\ref{fig:rho-cdw} for an intermediate regime of magnetic fields and compared to the case with no interactions.
We observe an CDW-induced increase in resistivity by a factor of $\sim 2 - 4$ in the range of $1.25 T < B < 3.75 T $ for our choice of parameters. 
This relative increase is robust to different parameter sets for the lattice model and mean-field parameters (e.g. values of the interaction U), where the exact range of the magnetic field is still affected by the size of the pockets due to the scaling properties discussed above.
Note that, upon approaching the quantum limit at $B_Q$, this factor slowly decreases with increasing $B$.
We can intuitively understand this behaviour remembering that the fraction of gapped levels is constant at low magnetic fields and must change at high magnetic fields.

\begin{figure}[ht]
    \centering
    \includegraphics[width=0.47\textwidth]{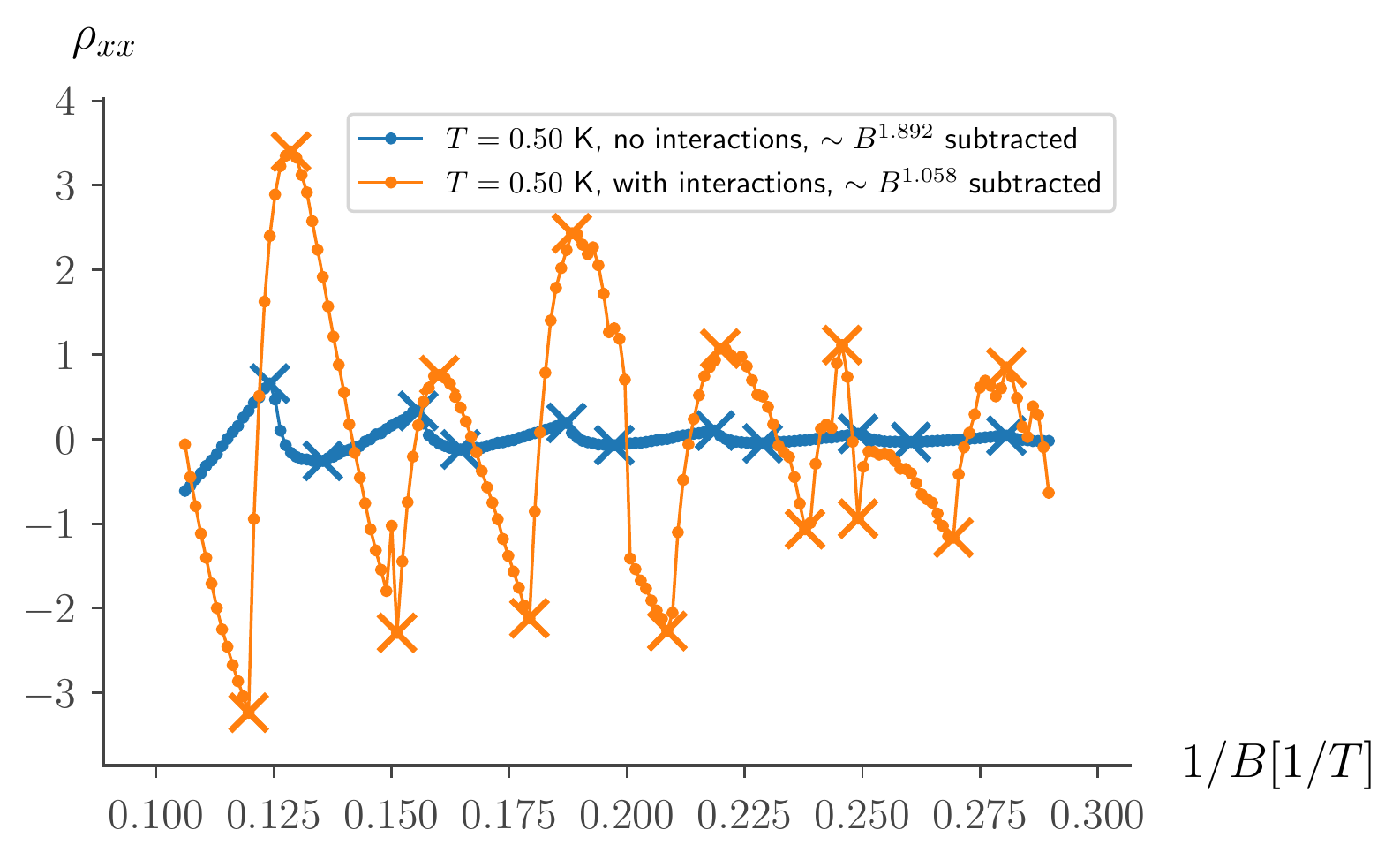}
    \includegraphics[width=0.47\textwidth]{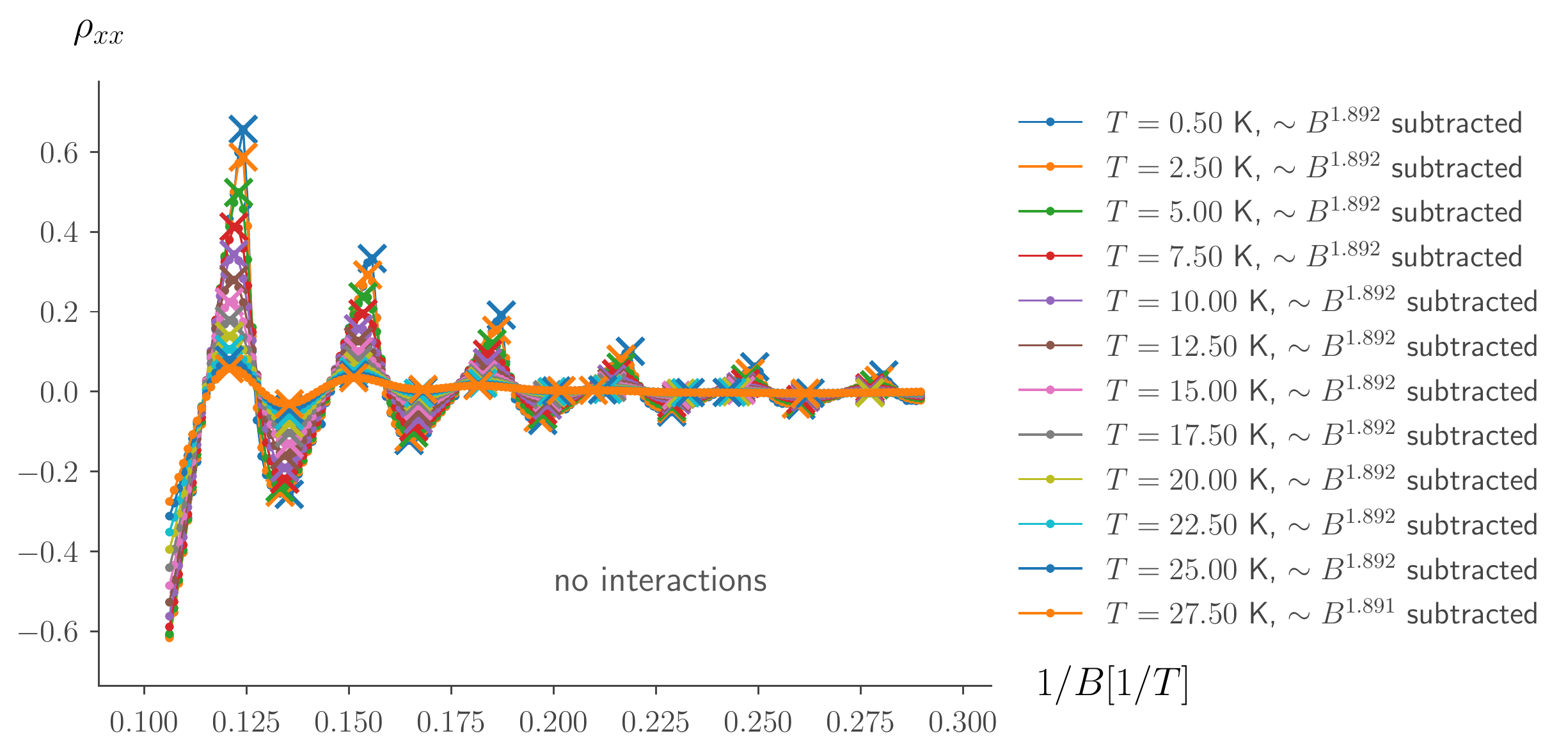}
    \includegraphics[width=0.47\textwidth]{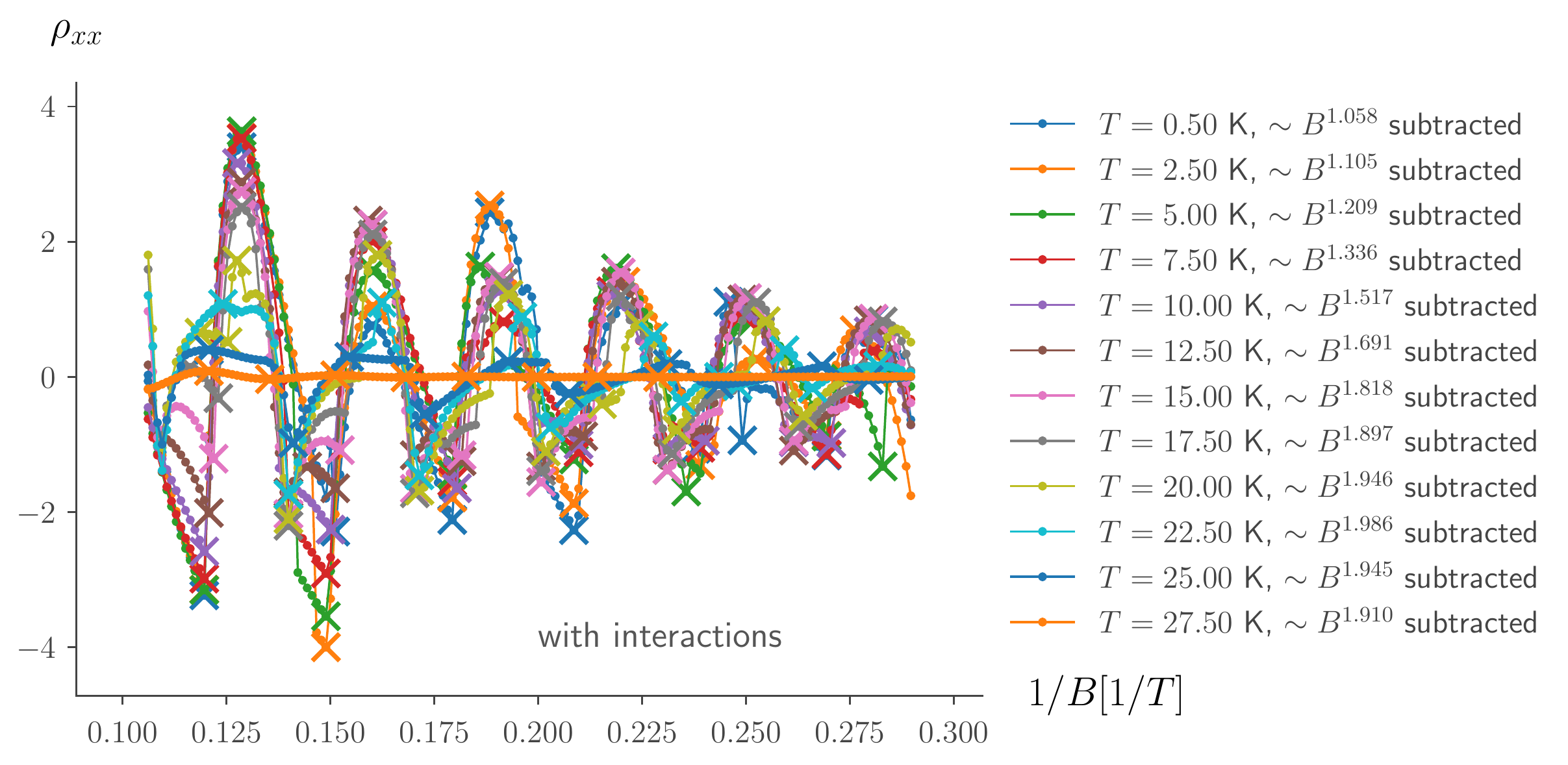}
    \caption{Quantum oscillations compared with and without interaction induced CDW.
    The power law background is subtracted as discussed in the main text.
    In the top panel oscillations with and without interactions are compared at $T = 0.5K$. The middle panel shows how oscillations without interactions evolve with different temperatures. The bottom panel displays the temperature evolution when including interactions.
}
    \label{fig:oscillations-cdw}
\end{figure}
\subsection{Quantum oscillations}
In the large magnetic field regime quantum oscillations from the discrete nature of the Landau levels appear. In Fig.\ref{fig:oscillations-cdw} we show quantum oscillations from low ($T=0.5K$) to high ($T=27.5K$) temperatures, where ``high'' temperature has to be understood in comparison to the critical temperature of the CDW (here $\simeq 25K$).

The figure shows only the oscillatory part of $\rho$, where the power law behaviour was subtracted,
vs. inverse magnetic field.
The power law is obtained by a numerical least-squares fit in log-log space. As expected, the frequency of oscillations is directly proportional to the area of the electron and hole pockets\cite{obrien_magnetic-klein-weyl2_2016}. 
Moreover, we conclude that the CDW formation significantly increases the amplitudes of the quantum oscillations in resistivity. Note, experimental observations of the quantum oscillations usually consider the conductivity (the Shubnikov-De Haas effect), which, being the inverse resistivity, would be decreased by the CDW. 
\begin{figure}[ht]
    \centering
    \includegraphics[width=0.48\textwidth]{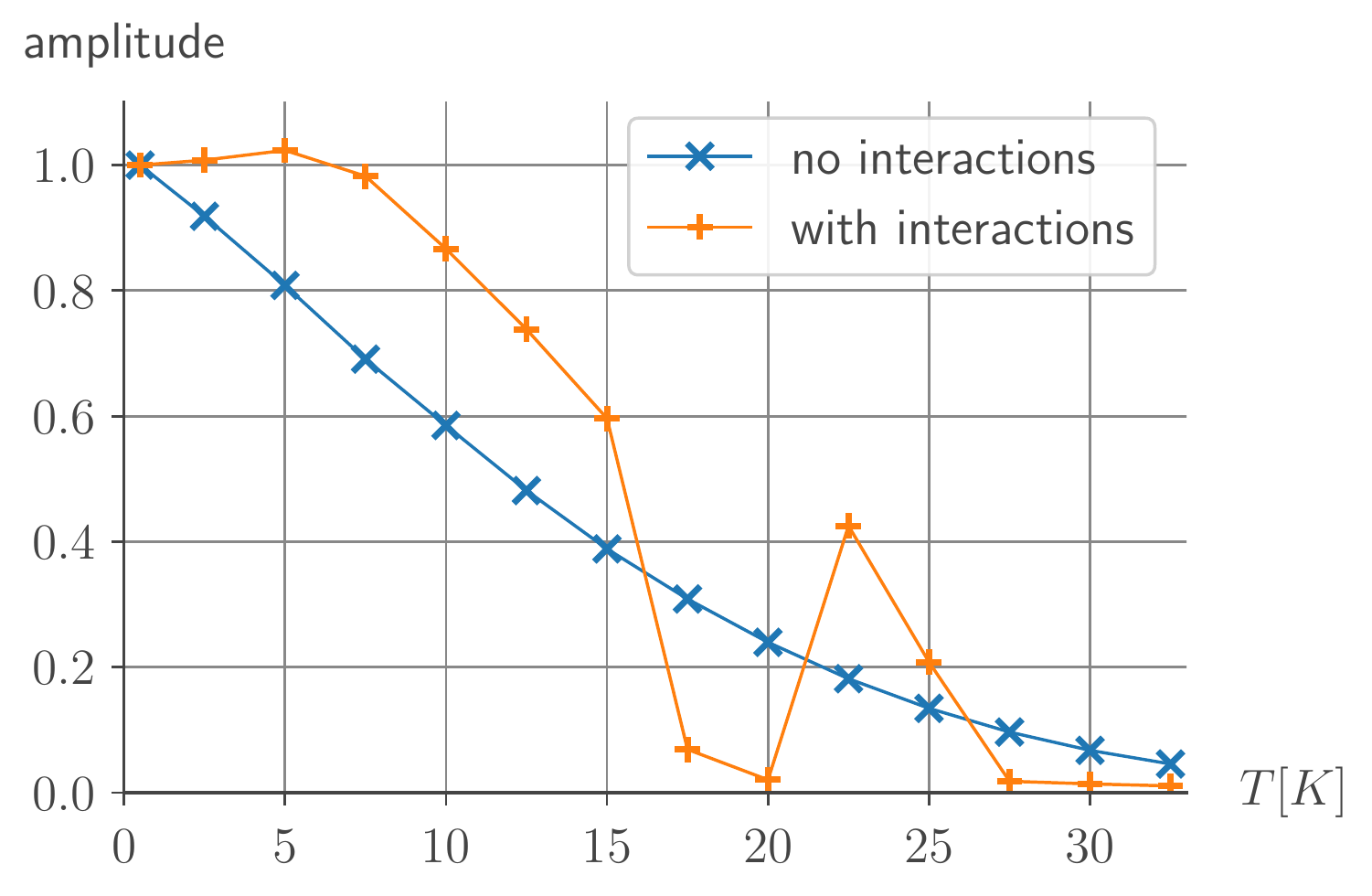}
    \caption{Amplitudes of the magnetoresistance oscillations vs. temperature. Amplitudes are normalized to $1$ at $T=0.5K$ (the smallest temperature value considered) for the two cases independently. The numerical results are shown as crosses, the solid lines are a guide to the eye.}
    \label{fig:oscillations-cdw-amplitudes}
\end{figure}

A main result of the present study concerns the impact of the CDW formation on the temperature dependence of the quantum oscillation amplitudes. 
We determine the amplitudes from the oscillatory part of $\rho$ as shown in Fig. \ref{fig:oscillations-cdw}.
We define the amplitude as the difference of a consecutive local maximum and minimum below a fixed magnetic field value,
and we checked that the qualitative behaviour is independent of the position at which the amplitude is determined.
In Fig.\ref{fig:oscillations-cdw-amplitudes} we compare the decay of the amplitude for increasing temperature for a representative case with and without CDW. 
While the case without CDW (blue cross) shows the usual monotonic decrease similar to the Lifshitz-Kosevich temperature dependence \cite{shoenberg_magnetic_1984}, the system with CDW clearly deviates from this universal behavior. 
More importantly, we observe a clear plateau and then an increase of the amplitude for increasing $T$ at very low temperatures.
This result is in clear contrast to the monotonic decrease of the Lifshitz-Kosevich temperature dependence.
We expect both signatures  to be clearly visible in experiments.

We stress that the oscillations in the CDW case are less regular both in temperature and in magnetic field, as shown in the bottom panel of Fig. \ref{fig:oscillations-cdw}.
This behaviour can be understood considering the cascade of CDW transitions discussed in Ref. \onlinecite{trescher_charge_2017}.
In the non-interacting case, there is only one way of gapping out Landau levels: by increasing the magnetic field and thereby the Landau level spacing, pushing Landau levels above the Fermi energy.
In the interacting case, however, for each Landau level there is a competition between two mechanisms: gapping the Landau level by a CDW transition or by the simple increase of the Landau level spacing similar to the non-interacting case.
Which mechanism is responsible for gapping out a specific Landau level depends on the magnetic field and on the temperature. This dependence is different for each Landau level\cite{trescher_charge_2017}.
Together, this leads to the complex behaviour of the oscillations in the interacting case, as seen in Fig.~\ref{fig:oscillations-cdw} (bottom panel).

The highest critical CDW temperature for any Landau level is $\simeq 25K$ in our exemplary model. Below this temperature, the cascade of CDW transitions affects the oscillations, and the oscillation amplitude is not as clearly defined as in the non-interacting case.
Following our fixed protocol of determining the oscillation amplitudes, we obtain the results shown in Fig. \ref{fig:oscillations-cdw-amplitudes}.
These amplitudes correspond to the first set of consecutive minima and maxima, i.e. at the highest magnetic field as shown in Fig. \ref{fig:oscillations-cdw}, where local extrema are marked by crosses.

\section{Discussion}
We calculated the magnetoresistance  of a type-II Weyl semimetal without and with a magnetic field induced CDW.
Earlier, the unsaturated and (nearly) quadratic magnetoresistance of several compounds was attributed to the compensation of electron and hole pockets, based on semiclassical computations.
Here, we confirmed this result for non-interacting type-II Weyl semimetals using a microscopic model and linear response theory.
Further, we showed that a critical field strength $B_Q$ exists, at which the quadratic scaling breaks down even for perfect compensation.
We discussed how this field strength, which corresponds to the quantum limit, depends on the materials parameters.

We then focussed on the effects of interactions, which lead to a CDW transition as reported in Ref. \onlinecite{trescher_charge_2017}.
In terms of magnetoresistance this CDW transition leads to numerous effects:
At low fields, the resistivity (and hence the magnetoresistance) is increased significantly and at high fields the resistivity approaches again the values of the non-interacting case.
Most importantly, the temperature dependence of the resistivity oscillations is in contrast to the universal Lifshitz-Kosevich behaviour.
Therefore, magnetoresistance measurements are an appropriate tool to look for signatures of the conjectured CDW phase in type-II Weyl semimetals.

We note that the number of Landau levels crossing the Fermi level is determined by the magnetic field component $B^\perp$ projected onto the tilt direction.
Hence, a magnetoresistance from Landau level formation of the form $(B^\perp)^2$ suggests a $\cos^2 \theta$ angular dependence similar to measurements on WP$_2$ \cite{kumar_extremely_2017}.

At the time of this writing, it seems likely that all materials showing unsaturated magnetoresistance may indeed have perfectly compensated electron and hole pockets. 
This has been consensus for WTe$_2$.
Another material, WP$_2$, likewise displaying an extreme magnetoresistance but with a more controversial provenance, has recently been argued to be particle-hole compensated as well\cite{razzoli_stable_2018}. For WTe$_2$ it indeed appears, that the extreme magnetoresistance is sensitively depending on fine-tuned particle hole symmetry\cite{gong_non-stoichiometry_2017,caputo_dynamics_2018}.
  
We also note that an extreme magnetoresistance has been identified in the topologically trivial material LaAs\cite{yang_extreme_2017}. 
In our setup it is quite clear that the topology of the Weyl node is not the root  of this effect---nevertheless, the band structure of a type-II Weyl node is perfectly primed to allow this effect.
Via the emergent nesting in a strong magnetic field, we argue that the appearance of a CDW is almost unavoidable.

Intriguing prospectives for future research include taking into account
interactions within the $n=0$ chiral Landau level, to consider finite frequency optical conductivity and to investigate the effect of doping away from particle-hole compensation. The latter generalization would include the relaxation of several approximations made and justified in Ref. \onlinecite{trescher_charge_2017}, including the simple contact interaction and the restriction of the interaction to Landau level branches with equal Landau level index. 

Finally, beyond Weyl semimetals it will be worthwhile to extend the general idea that field-induced density wave transitions can strongly affect the amplitude scaling of quantum oscillation measurements\cite{pezzini_unconventional_2017,shishido_anomalous_2018}, and hence interpretation of the corresponding data. 

\begin{acknowledgments}
We acknowledge useful conversations and related collaborations with  Bj\"orn Sbierski, Marcus St\aa lhammar and Masafumi Udagawa. We thank C. Felser, J. Gooth and A. Pronin for discussions.
This work was supported by  Emmy Noether program (BE 5233/1-1) of the Deutsche Forschungsgemeinschaft, the Swedish research council (VR) and the Wallenberg Academy Fellows program of the Knut and Alice Wallenberg Foundation.  
\end{acknowledgments}

\appendix
\onecolumngrid

\section{Matrix Elements}
\label{sec:matrixelements}
After the publication of Ref. \onlinecite{trescher_charge_2017} we gained some more insight in the structure of the interaction matrix elements, also generalizing to arbitrary combinations of Landau level indices $n_{1,2,3,4}$, as defined in the main text and the supplementary material of Ref. \onlinecite{trescher_charge_2017}.
Recall the matrix elements formula
 \begin{align}
    M_{n_1,n_2,n_3,n_4}(\vec{q}) &=
        U(\vec{q})
        \underbrace{ J_{n_4,n_1}(q_x, q_y) J_{n_3,n_2}(-q_x, -q_y) }_{F(q_x, q_y)} 
\end{align}
\begin{align}
    J_{m,n}(q_x, q_y) &= \sqrt{\frac{n!}{m!}} e^{-|q|^2/4} \left( \frac{q_x + i q_y}{\sqrt{2}} \right)^{m-n} L^{m-n}_n \left( \frac{|q|^2}{2}\right)  \quad \textrm{with } |q|^2 = q_x^2 + q_y^2 \quad \textrm{for } \; m < n \\
    J_{m,n}\left(q_x, q_y\right) &= J_{n,m}^{*}\left( -q_x, -q_y \right) ,
\end{align}
that we want to evaluate at finite $k_z$ but $q_x=q_y=0$.
The following analytic results are obtained in the basis of the $c_A, c_B$ operators, therefore a basis transformation introducing a dependence on $n, k_z$ and $B$ needs still to be applied.

\subsection{Hartree terms}
In $J_{m,n}$ there appears a factor of $(q_x \pm i q_y)^{m-n}$, as the matrix elements for the Hartree terms are evaluated at $q_x = q_y = 0$, this factor cancels any contribution for $m\neq n$ and the only contribution is for $m=n$, giving a factor of $1$, as $L_n^{0}(0)=1$.
Due to the shift of $c_{\alpha=B,n} \rightarrow c_{B,n+1}$, the factor $J$ corresponding to $c_{B,n_i}$ depends on $n_i -1$, leading to the full interaction matrix:
\begin{align}
\left[\begin{matrix}\delta_{n_{1} n_{4}} \delta_{n_{2} n_{3}} & \delta_{n_{1}, n_{4} - 1} \delta_{n_{2} n_{3}} & \delta_{n_{1} n_{4}} \delta_{n_{2}, n_{3} - 1} & \delta_{n_{1}, n_{4} - 1} \delta_{n_{2}, n_{3} - 1}\\\delta_{n_{1} n_{4}} \delta_{n_{3}, n_{2} - 1} & \delta_{n_{1}, n_{4} - 1} \delta_{n_{3}, n_{2} - 1} & \delta_{n_{1} n_{4}} \delta_{n_{2} - 1, n_{3} - 1} & \delta_{n_{1}, n_{4} - 1} \delta_{n_{2} - 1, n_{3} - 1}\\\delta_{n_{2} n_{3}} \delta_{n_{4}, n_{1} - 1} & \delta_{n_{2} n_{3}} \delta_{n_{1} - 1, n_{4} - 1} & \delta_{n_{2}, n_{3} - 1} \delta_{n_{4}, n_{1} - 1} & \delta_{n_{2}, n_{3} - 1} \delta_{n_{1} - 1, n_{4} - 1}\\\delta_{n_{3}, n_{2} - 1} \delta_{n_{4}, n_{1} - 1} & \delta_{n_{3}, n_{2} - 1} \delta_{n_{1} - 1, n_{4} - 1} & \delta_{n_{4}, n_{1} - 1} \delta_{n_{2} - 1, n_{3} - 1} & \delta_{n_{1} - 1, n_{4} - 1} \delta_{n_{2} - 1, n_{3} - 1}\end{matrix}\right]
    \label{}
\end{align}

\subsection{Fock terms}
To calculate the Fock terms of the mean-field expansion we can use the rewriting introduced in  Ref. \onlinecite{goerbig_competition_2004}.
Therefore we are interested in $\tilde{M}(\vec{q})$ at $\vec{q}=0$, which turns out to be the integral over $q_x, q_y$ of the $M$
Introducing the complex variable $z = q_x + i q_y$ this integral is equivalent to the integral over the complex plane, and the argument of the Laguerre polynomials becomes $|z|^2$.
For each factor of $J$ in the matrix element we get $z^\alpha$ with $\alpha = m-n$ being the second index of the associated Laguerre polynomial.
Further we distinguish $\alpha = n_4 - n_1$ and $\alpha' = n_3 - n_2$ and we use the polar notation $z = r e^{i\varphi}$.
Note that negative $\alpha$ corresponds to exchanging the two indices of the Laguerre polynomial and additionally introduces a complex conjugation.
Now two cases can occur:
\begin{enumerate}
    \item $\alpha = -\alpha'$: this leads to an integral of the form
\begin{align}
    \int_C \d{z} z^\alpha {z^*}^{-\alpha}  L_n^\alpha L_{n'}^\alpha e^{-z^2}
    =& \int_C \d{z} |z|^{2\alpha} L_n^\alpha L_{n'}^\alpha e^{-|z|^2}
    \label{}
\end{align}
which is exactly the orthogonality relation of the associated Laguerre polynomials, leading to $\delta_{n,n'}$, with $n = \max{n1,n4}$ and $n'=\max{n2,n3}$ with the appropriate $\pm 1$ shift for the $B$ operators.

\item
    $\alpha \neq -\alpha'$: We introduce $\beta$ as $\alpha' = -\alpha - \beta$ and manifestly nonzero $\beta$
The integrand now includes a factor of $|z|^{2\alpha} z^\beta$ (or ${z^*}^\beta$).
The part $|z|^{2\alpha} L_n^\alpha L_{n'}^\alpha e^{-|z|^2}$ of the integrand only depends on $|z|$, i.e. is independent of the complex phase, and consequently the integral over the complex plane can be split into integrations over $r$ and $\varphi$, and the integration $\int \d{\varphi} z^n$ is $0$.
As a negative value of $\alpha$ always is related to the complex conjugate of $J$ with positive $\alpha$, this case rules out any contribution of elements where the product of two Laguerre polynomials with different $\alpha$ indices occurs.
\end{enumerate}

Together this leads to the following matrix elements of $\tilde{M}$ (wrapping the $4\times 4$ matrix to two lines):
\begin{align}
\left[\begin{matrix}\delta_{- n_{1} + n_{4}, n_{2} - n_{3}} \delta_{\max\left(n_{1}, n_{4}\right), \max\left(n_{2}, n_{3}\right)} & \delta_{n_{2} - n_{3}, - n_{1} + n_{4} - 1} \delta_{\max\left(n_{1}, n_{4} - 1\right), \max\left(n_{2}, n_{3}\right)}\\\delta_{- n_{1} + n_{4}, n_{2} - n_{3} - 1} \delta_{\max\left(n_{1}, n_{4}\right), \max\left(n_{3}, n_{2} - 1\right)} & \delta_{- n_{1} + n_{4} - 1, n_{2} - n_{3} - 1} \delta_{\max\left(n_{1}, n_{4} - 1\right), \max\left(n_{3}, n_{2} - 1\right)}\\\delta_{n_{2} - n_{3}, - n_{1} + n_{4} + 1} \delta_{\max\left(n_{2}, n_{3}\right), \max\left(n_{4}, n_{1} - 1\right)} & \delta_{- n_{1} + n_{4}, n_{2} - n_{3}} \delta_{\max\left(n_{2}, n_{3}\right), \max\left(n_{1} - 1, n_{4} - 1\right)}\\\delta_{- n_{1} + n_{4} + 1, n_{2} - n_{3} - 1} \delta_{\max\left(n_{3}, n_{2} - 1\right), \max\left(n_{4}, n_{1} - 1\right)} & \delta_{- n_{1} + n_{4}, n_{2} - n_{3} - 1} \delta_{\max\left(n_{3}, n_{2} - 1\right), \max\left(n_{1} - 1, n_{4} - 1\right)}\end{matrix}\right. \nonumber \\
\left.\begin{matrix}\delta_{- n_{1} + n_{4}, n_{2} - n_{3} + 1} \delta_{\max\left(n_{1}, n_{4}\right), \max\left(n_{2}, n_{3} - 1\right)} & \delta_{- n_{1} + n_{4} - 1, n_{2} - n_{3} + 1} \delta_{\max\left(n_{1}, n_{4} - 1\right), \max\left(n_{2}, n_{3} - 1\right)}\\\delta_{- n_{1} + n_{4}, n_{2} - n_{3}} \delta_{\max\left(n_{1}, n_{4}\right), \max\left(n_{2} - 1, n_{3} - 1\right)} & \delta_{n_{2} - n_{3}, - n_{1} + n_{4} - 1} \delta_{\max\left(n_{1}, n_{4} - 1\right), \max\left(n_{2} - 1, n_{3} - 1\right)}\\\delta_{- n_{1} + n_{4} + 1, n_{2} - n_{3} + 1} \delta_{\max\left(n_{2}, n_{3} - 1\right), \max\left(n_{4}, n_{1} - 1\right)} & \delta_{- n_{1} + n_{4}, n_{2} - n_{3} + 1} \delta_{\max\left(n_{2}, n_{3} - 1\right), \max\left(n_{1} - 1, n_{4} - 1\right)}\\\delta_{n_{2} - n_{3}, - n_{1} + n_{4} + 1} \delta_{\max\left(n_{4}, n_{1} - 1\right), \max\left(n_{2} - 1, n_{3} - 1\right)} & \delta_{- n_{1} + n_{4}, n_{2} - n_{3}} \delta_{\max\left(n_{1} - 1, n_{4} - 1\right), \max\left(n_{2} - 1, n_{3} - 1\right)}\end{matrix}\right]
    \label{fock}
\end{align}

In most of the cases $n_1$ has to be equal (up to $\pm 1$) to $n_3$, while it is independent of $n_2, n_4$, which in turn are equal up to $\pm 1$ as well.
This is comparable to the situation of the Hartree results were $n_1,n_4$ were independent of $n_2, n_3$, but here with switched pairs.
One can easily verify this when $n_1, n_3$ are much larger (or smaller) than $n_2$, $n_4$, as the terms $ \delta_{\max\left(n_{1}, n_{4}\right), \max\left(n_{2}, n_{3}\right)} $ then can be simplified, regardless of some $\pm 1$.

\end{document}